\documentclass[12pt]{iopart}
\usepackage{amsfonts}
\usepackage{graphicx}
\usepackage{epsfig}
\usepackage{graphics}

\makeatletter
\newcommand{\row}[1]%
{\mathord{\buildrel{\lower3pt%
\hbox{$\scriptscriptstyle\rightarrow$}}\over #1}}
\newcommand{\col}[1]{{#1^{\raisebox{2pt}[\height]%
{$\scriptstyle\downarrow$}}}}
\newcommand{\dyadic}[1]{\mathord{\dyadic@rrow{#1}}}
\newcommand{\dyadic@rrow}[1]{
\begin{picture}(12,12)(-1,0)
\put(-2,12){\makebox(0,0)[t]{$\scriptscriptstyle\downarrow$}}
\put(-2,12){\makebox(0,0)[l]{$\scriptscriptstyle\longrightarrow$}}
\put(5,0){\makebox(0,0)[b]{$#1$}}
\end{picture}
}

\newcommand{\bra}[1]{\bigl\langle #1 \bigr|}
\newcommand{\ket}[1]{\bigl| #1 \bigr\rangle}
\newcommand{\expect}[1]{\left\langle #1 \right\rangle}
\newcommand{\PRA}[3]{Phys.\ Rev.\ A  \textbf{#1}, #2 (#3)}
\newcommand{\plA}[3]{Phys.\ Lett.\ \textbf{A#1}, #2 (#3)}

\begin{document}

\title {A paradigm for entanglement and information swapping of two qubits}

\author{N. Metwally$^a$ and  M. Abdel-Aty$^b$  }
\address{$^a$Mathematics Department, Faculty of Science, South Valley
University,aswan, Egypt.\\
 $^b$Mathematics Department, Faculty of Science, South Valley
University,aswan, Egypt.}

\begin{abstract}
Based on a general unitary operation of two-qubit system, we
investigate the dynamics of entanglement, entropy and purity.
Distinctive features in the different initial states settings of
an interacting bipartite system are explored. The striking effects
of initial product states on entanglement shows that zeros
entanglement can be obtained during the development of the
interaction.

\end{abstract}

\maketitle
\section{Introduction}

To test fundamental quantum concepts and implement various
potential applications, including sensitive detection and quantum
information processing, the engineering of quantum states has
attracted considerable attention in recent years \cite{mon96}. On
the other hand, entanglement takes an important part in
manipulating  with information, storing, as  in quantum coding
\cite{ben,Metwally}, sending, as in  teleportation \cite{bennt}.
So, generating  and quantifying the entanglement is one of the
most important tasks in quantum information context. In quantum
computer, qubits are the arithmetical units \cite{Deutch}, which
makes a new type of dynamics able to solve different problems
faster than classical computer. It  is desirable  then  to discuss
the behavior of these qubits when a switch on/of  the computer is
taken place.

Investigating the dynamics of the purity of quantum system is one
of the most important topics in the context of quantum information
theory \cite{ Rubn, Tomz}. It  can be used as a measure of the
degree of entanglement for mixed state \cite{Ade, Ser}.
 Bagan et. al. have been estimated the purity of  a large
number N of copies of a qubit state \cite{Ball}.
 Our aim in the present article is twofold: The first is
to discuss the dynamics of two separable qubits under the effect
of the unitary operation. The second is to treat the phenomenon of
swapping information from a sender (Alice) to a receiver (Bob). We
suggest that Alice has a  completely polarized qubit in
x-direction, while Bob's qubit takes any form. We show that the
purity of one qubit can purified on the expanse of the other one.

The rest of the paper is arranged as follows: Section $2$, is
devoted to describe the general two-qubit pairs by means of its
Bloch vectors and a cross dyadic, which is defined by $3\otimes 3$
matrix. In Section $3$, we obtain the density operator for the
individual subsystems, which can be used to read out how much
information lost or gained. Several examples are given in Section
$4$, in which the initial states are assumed to be product states.
In section $5$, we consider a class for entangled states. The
paper ends with a conclusion in Sec.$6$.

\section{Two qubit pairs}

Assume that the user Alice and Bob are given a state of a two
qubit pairs to perform any quantum information task. The general
form of the 2-qubit state is
\begin{equation}
\rho_{ab}=\frac{1}{4}(1+\row{s}\cdot\col{\sigma}+\row{t}\cdot\col\tau+\row\sigma
\cdot\dyadic{C}\cdot\col\tau),
\end{equation}
where $\row\sigma$ and $\row\tau$ are the Pauli's spin vector of
the first and the second qubits respectively \cite{Englert}. The
statistical operator for the individual quibits are specified by
their Bloch vectors, $\row{s}=\expect{\row\sigma}$ and
$\row{t}=\expect{\row\tau}$. The cross dyadic $\dyadic{C}$ is
represented  by a $3\times 3$ matrix. it describes the correlation
between the first qubit,
$\rho_a=\tr_{b}\{\rho_{ab}\}=\frac{1}{2}(1+\row{s}\cdot\col\sigma)$
and the second qubit
$\rho_b=\tr_{a}\{\rho_{ab}\}=\frac{1}{2}(1+\row{t}\cdot\col\tau)$.
The Bloch vectors and the cross dyadic are given by
\begin{eqnarray}
\row{s}&=&(s_x,s_y,s_z) ,\quad \row{t}=(t_x,t_y.t_z),\quad
\mbox{and}~ \dyadic{C}= \left(
\begin{array}{ccc}
c_{xx}&c_{xy}&c_{xz}\\
c_{yx}&c_{yy}&c_{yz}\\
c_{xx}&c_{zy}&c_{zz}
\end{array}
\right)
\end{eqnarray}

 Now, we  suppose that the interacting  Hamiltonian of the two
qubits is given by
\begin{equation}
\mathcal{H}=\sum_{i=1}^3{\alpha_i\sigma_i\otimes
I}+\sum_{j=1}^3{\beta_j
I\otimes\tau_i}+\sum_{i,j=1}^3{\alpha_{ij}\sigma_i\otimes\tau_j}
\end{equation}
The first two terms represent the free evolution of the individual
two qubits, while the third term represents the evolution of the
correlation part. It is described by a dyadic consists of $9$
elements. By using a suitable  local unitary transformation, we
can  always choose a reference system in which this dyadic can be
reduced to a diagonal one. In this case we have only $3$ elements.
In our treatment we shall consider  the effective Hamiltonian
only. In the computational basis the unitary operator \cite{Kraus}
is given by
\begin{eqnarray}\label{eq:un}
\mathcal{U}&=&\Gamma_1(\ket{00}\bra{00}+\ket{11}\bra{11})+
\Gamma_2(\ket{01}\bra{01}+\ket{10}\bra{10})
\nonumber\\
&+&\Gamma_3(\ket{10}\bra{01}+\ket{01}\bra{10})+
\Gamma_4(\ket{11}\bra{00}+\ket{00}\bra{11})
\end{eqnarray}
where,
\begin{eqnarray}
\Gamma_1&=&e^{-i\alpha_3t}\cos t(\alpha_1-\alpha_2),\quad\quad
\Gamma_2=e^{i\alpha_3t}\cos t(\alpha_1+\alpha_2),
\nonumber\\
\Gamma_3&=&-ie^{-i\alpha_3t}\sin t(\alpha_1+\alpha_2),\quad
\Gamma_4=ie^{i\alpha_3t}\sin t(\alpha_1-\alpha_2),
\end{eqnarray}
This is canonical unitary operator \cite{Ramos,Wang}. Under the
canonical evolution  the density matrix of the 2-qubit state
transforms to
\begin{eqnarray}\label{eq:new}
\rho^{\mathrm{new}}&=&\mathcal{U}\frac{1}{2}(1+\row{s}\cdot\col{\sigma})
\frac{1}{2}(1+\row{t}\cdot\col{\tau})\mathcal{U}^{\dagger}
\nonumber\\
&=&\frac{1}{4}(1+\row{s_{\mathrm{new}}}\cdot\col\sigma
+\row{t_{\mathrm{new}}}\cdot\col\tau+\row\sigma\cdot\dyadic{C}_\mathrm{new}\cdot\col\tau),
\end{eqnarray}
where the new Bloch vectors and the dyadic are given by,
\begin{eqnarray}
\row{s_{\mathrm{new}}}=(s_1,s_2,s_3),\quad
\row{t_{\mathrm{new}}}=(t_1,t_2,t_3), \quad
\dyadic{C}_{\mathrm{new}}= \left(
\begin{array}{ccc}
c_{11}&c_{12}&c_{13}\\
c_{21}&c_{22}&c_{23}\\
c_{31}&c_{32}&c_{33},
\end{array}
\right)
\end{eqnarray}
with
\begin{eqnarray}
s_1&=&\cos2\alpha_3(s_x\cos2\alpha_1+c_{zy}\sin 2\alpha_1)+ \sin
2\alpha_3 (t_x \sin 2\alpha_1-c_{yz}\cos 2\alpha_1),
 \nonumber\\
s_2&=&(1-2\cos^2\alpha_3)(s_y \cos2\alpha_2-c_{zx}\sin 2\alpha_2)
-\sin 2\alpha_3(t_y\sin2 \alpha_2+c_{xz} \cos 2\alpha_2),
\nonumber\\
s_3&=&s_z[cos^2(\alpha_1+\alpha_2)+cos^2(\alpha_1-\alpha_2)-1]
+t_z[cos^2(\alpha_1+\alpha_2)-cos^2(\alpha_1-\alpha_2)]
\nonumber\\
&-&\frac{1}{2}(c_{xy}+c_{yx})\left[\sin 2(\alpha_1-\alpha_2) +\sin
2(\alpha_1+\alpha_2)\right],
\end{eqnarray}

\begin{eqnarray}
t_1&=& \sin 2\alpha_3(s_x\sin 2\alpha_1-c_{zy}\cos 2\alpha_1)
-(1-2\cos^2\alpha_3)(t_x\cos 2\alpha_1+C_{yz}\sin 2\alpha_1),
\nonumber\\
t_2&=&-\sin 2\alpha_3(s_y\sin 2\alpha_2+c_{zx}\cos 2\alpha_2)+
(1-2\cos^2\alpha_3)(t_y\cos_2 \alpha_2-c_{xz}\sin 2\alpha_2),
\nonumber\\
t_3&=&s_z\left(cos^2(\alpha_1-\alpha_2)-\cos^2(\alpha_1+\alpha_2)\right)
+t_z\left(cos^2(\alpha_1-\alpha_2)+\cos^2(\alpha_1+\alpha_2)-1\right)
\nonumber\\
&+&\frac{1}{2}(c_{xy}+c_{yx})\bigl[\sin 2(\alpha_1-\alpha_2)- \sin
2(\alpha_1+\alpha_2)\bigr],
\end{eqnarray}
and
\begin{eqnarray}
c_{11}&=&c_{xx}, \nonumber\\
c_{12}&=&-\frac{1}{2}(s_z+t_z)\left[\sin 2(\alpha_1-\alpha_2)-sin
2(\alpha_1+\alpha_2)\right],
\nonumber\\
&-&(c_{xy}+c_{yx})\left[\cos^2(\alpha_1-\alpha_2)+\cos^2(\alpha_1+\alpha_2)-1\right],
\nonumber\\
c_{13}&=&(s_y-c_{zx})\sin 2\alpha_3\sin
2\alpha_2-(1-2\cos^2\alpha_3)(t_y\sin 2\alpha_2+c_{xz}\cos
2\alpha_2),
 \nonumber\\
 \nonumber\\
c_{21}&=&-\frac{1}{2}(s_z+t_z)\left[\sin 2(\alpha_1-\alpha_2)-\sin
2(\alpha_1+\alpha_2)\right]
\nonumber\\
&-&(c_{xy}+c_{yx})\left[\cos^2(\alpha_1-\alpha_2)-\cos^2(\alpha_1+\alpha_2)\right],
\nonumber\\
c_{22}&=&c_{yy} \nonumber\\
c_{23}&=&-\sin 2\alpha_3(s_x\cos 2\alpha_1-c_{zy}\sin 2\alpha_1)
-(1-2\cos^2 \alpha_3)(t_x \sin 2\alpha_1-c_{yz}\cos 2\alpha_1),
\nonumber\\
\nonumber\\
 c_{31}&=&-\sin2\alpha_3(t_y\cos 2\alpha_2-c_{xz}\sin
2\alpha_2) -(1-2\cos^2\alpha_3)(s_y\sin 2\alpha_2+c_{zx}\cos
2\alpha_2),
\nonumber\\
c_{32}&=&\sin2\alpha_3(t_x\cos 2\alpha_1-c_{yz}\sin 2\alpha_1)
-(1-2\cos^2\alpha_3)(s_x\sin 2\alpha_1-c_{zy}\cos 2\alpha_1)
\nonumber\\
c_{33}&=&c_{zz}.
\end{eqnarray}
\section{The dynamics of the individual qubits}
In this section, we investigate how the density matrices of the
two qubits $\rho_a$ and $\rho_b$ evolve. Tracing out the second
qubit (Bob's qubit), we obtain the first qubit (Alice's qubit),as
\begin{equation}
\tilde{\rho_a}=\frac{1}{2}(1+\tilde {\row{ s}}\cdot\col\sigma),
\end{equation}
where
\begin{eqnarray}
\tilde{s_1}&=& \cos 2t\alpha_3(s_x
\cos2t\alpha_1+c_{zy}\sin2\alpha_1)+\sin 2t\alpha_3(t_x\sin
2t\alpha_1-c_{yz}\cos 2t\alpha_1),
 \nonumber\\
\tilde {s_2}&=&\cos 2t\alpha_3(c_{zx}\sin 2t\alpha_2-s_y\cos
2t\alpha_2)-\sin 2t\alpha_3(t_y\sin 2\alpha_2+c_{xz}\cos
2t\alpha_2),
\nonumber\\
\tilde
{s_3}&=&\frac{1}{2}(s_z+t_z)(\cos2t(\alpha_1+\alpha_2)+\cos2t(\alpha_1-\alpha_2))
\nonumber\\
&& -\frac{1}{2}(c_{xy}+c_{yx})(\sin
2t(\alpha_1+\alpha_2)+\sin2t(\alpha_1-\alpha_2),
\nonumber\\
\end{eqnarray}
Similarly, if we trace  out the first qubit we obtain  the density
matrix for the second qubit. In this context, we can write
$\rho_b$ in the following form,
\begin{equation}
\tilde \rho_b=\frac{1}{2}(1+\tilde{\row{t}}\cdot\col\tau),
\end{equation}
where,
\begin{eqnarray}
\tilde{t_1}&=&\sin 2t\alpha_3(s_x
\sin2t\alpha_1-c_{zy}\cos2t\alpha_1)+\cos 2t\alpha_3(t_x\cos
2t\alpha_1+c_{yz}\sin 2t\alpha_1),
\nonumber\\
\tilde{t_2}&=&-\sin2t\alpha_3(s_y\sin2t\alpha_2+c_{zx}\cos2t\alpha_2)-
\cos2t\alpha_3(t_y\cos2t\alpha_2+c_{xz}\sin2t\alpha_2),
\nonumber\\
\tilde{t_3}&=&\frac{1}{2}(s_z+t_z)[\cos2t(\alpha_1+\alpha_2)+\cos2t(\alpha_1-\alpha_2)]
\nonumber\\
&&+\frac{1}{2}(c_{xy}+c_{yx})[\sin2t(\alpha_1+\alpha_2)-\sin2t(\alpha_1-\alpha_2)],
\nonumber\\
\end{eqnarray}
Now, we have all the details to study the evolution of any
physical property of the total system which is represented by
$\rho_{ab}$ and its individual systems $\rho_a$ and $\rho_b$. For
example to quantify the  amount of entanglement contained in the
entangled states, we shall use   a measurement introduced  by K.
Zyczkowski \cite{Peres,Hor,Zyc}. This measure states that if the
eigenvalues of the partial transpose are given by $\lambda_j,
j=1,2,3,4$, then the degree of entanglement, DOE is defined by
\begin{equation}
DOE=\sum_{i=j}^{4}{|\lambda_j|-1}
\end{equation}

Also, we can study the evolution of the purity of the individual
subsystems which is given by,
\begin{equation}\label{purity}
\eta_i=tr{\rho_i^2},
\end{equation}
where $i=a,b$ (see for example \cite{Igor, Hari}). Further,the
amount of information contained in a density operator $\rho$ is
defined by the Von Neummann entropy
$\mathcal{S}=-\sum_j\lambda_i\ln{\lambda_i}$, where $\lambda_i$
are the eigenvalues of the density operator \cite{Garry}.
 If we start with an entangled
initial state $\rho_{ab}$, then we search for the survival amount
of entanglement under the effect of the Hamiltonian. In what
follows we shall consider  two classes: The first for product
states and the second for entangled pure state.
\section{Product Class }
{\it The first class:}
 In  this section, we consider the
evolution of the two qubits, where the  initial  two states are
given by
\begin{equation}\label{Eq:init}
\rho_a=\frac{1}{2}(1+s_x\sigma_x)~\quad \mbox{and}~
\rho_b=\frac{1}{2}I.
\end{equation}
Under the unitary transformation (\ref{eq:un}), the product state
, $\rho_a\otimes\rho_b$  of  the two qubits (\ref{Eq:init})
evolves to its final state with  density operator,
\begin{eqnarray}
\rho_{ab}&=&\frac{1}{4}(1+
s_x\cos(2t\alpha_1)\cos(2t\alpha_3)\sigma_x+s_x\cos(2t\alpha_1)\sin(2t\alpha_3)\tau_x)
\nonumber\\
&&-s_x\cos 2t\alpha_1\sin\
2t\alpha_3\sigma_y\tau_z-s_x(1-2\cos^2t\alpha_3) \sin
2t\alpha_1\sigma_z\tau_y )
\end{eqnarray}
 In Fig.$(1a)$, we plot the  degree of entanglement as a function of the scaled
time. In these calculations we assume that Alice and Bob perform
the same local unitary operation. Also the unitary operator is
defined where $\alpha_1=\alpha_2=\alpha_3=\frac{\pi}{6}$ and
$s_x=1$. It is clear that  as soon as the interaction is switched
on, an entangled state is generated. For any value of identical
$\alpha_i$, this  state swap from the entangled into separable
states. But  as $\alpha_i$ decreases, the robust of remaining in
entangled or separable state is decreased.

Since, we are interested in the behavior of the individual qubits
through the evolution, we calculate the density operator for each
qubit. Consequently,
\begin{equation}
\tilde \rho_1=\frac{1}{2}(1+\cos 2
t\alpha_1\cos2t\alpha_3\sigma_x),\quad   \tilde
\rho_2=\frac{1}{2}\bigl(1+\sin 2
t\alpha_1\sin2t\alpha_3\tau_x\bigr).
\end{equation}
\begin{figure}
  \begin{center}
  \includegraphics[width=25pc,height=10pc]{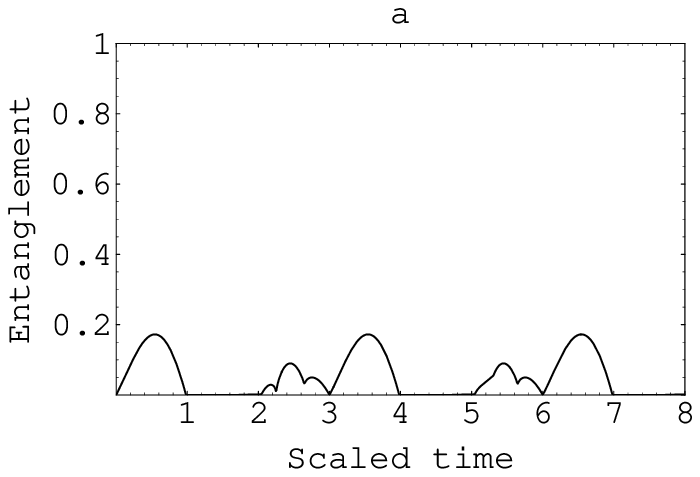}
 \includegraphics[width=15pc,height=9pc]{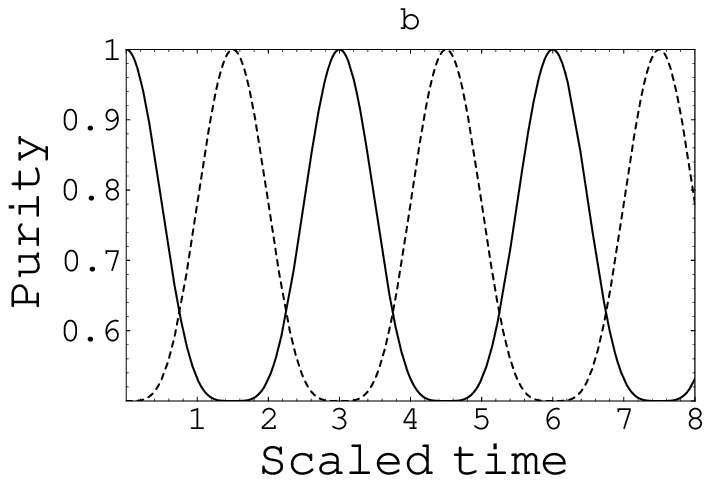}
 \includegraphics[width=15pc,height=9pc]{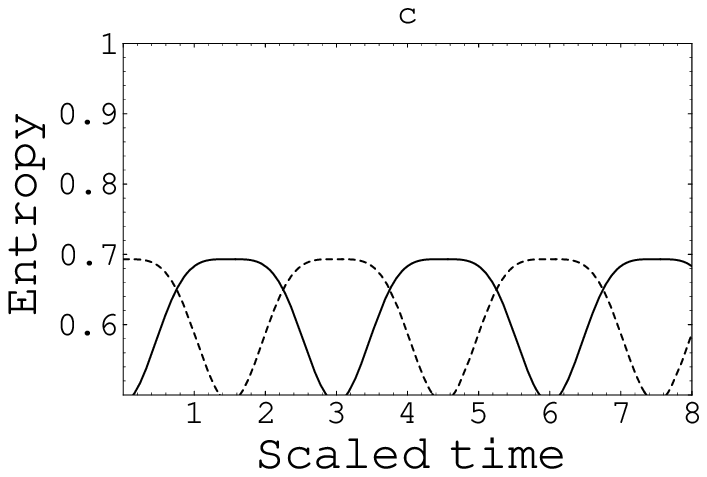}
  \caption{(a) shows the degree of entanglement,(b), shows the purity for the output
  states
  $\tilde\rho_a$(solid-line) and  $\tilde\rho_b$ (the dot-line).
  (c), shows the entropy for $\tilde\rho_a$(solid-line) and  $\tilde\rho_b$ (the dot-line).
The parameters are,  $\alpha_i=\frac{\pi}{6}$; $s_x=1, s_y=s_z=0$
and $t_x=t_y=t_z=0$.}
  \end{center}
\end{figure}
For these new states of the two qubits, we investigate how the
purity  and entropy evolve. In Fig.$(1b)$, we consider the purity
as a function of the scaled time, for  the  same parameters as in
Fig.$(1)$. From this figure we  see how the purity is completely
swapped from one pair to another. This phenomena is very important
in quantum information context. It is possible to Alice to send
all the information  she has.  We assume that Alice wants to
distribute a quantum key to Bob for performing a quantum
cryptography. She will code the secure key in her qubit by
applying any coding protocol. Then she agrees with Bob on the
unitary operators which they have to do. Fig.$(1c)$, shows the
behavior of the entropy for each qubit. It is clear that, the
phenomena of swapping  is satisfied for the entropy, where as the
entropy of one qubit vanish the purity of the second qubit is
maximum \cite{Metwally}.
\begin{figure}
  \begin{center}
  \includegraphics[width=25pc,height=10pc]{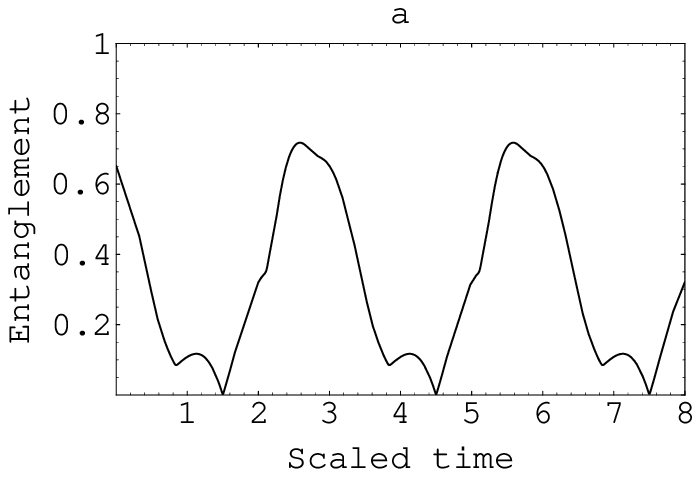}
 \includegraphics[width=15pc,height=9pc]{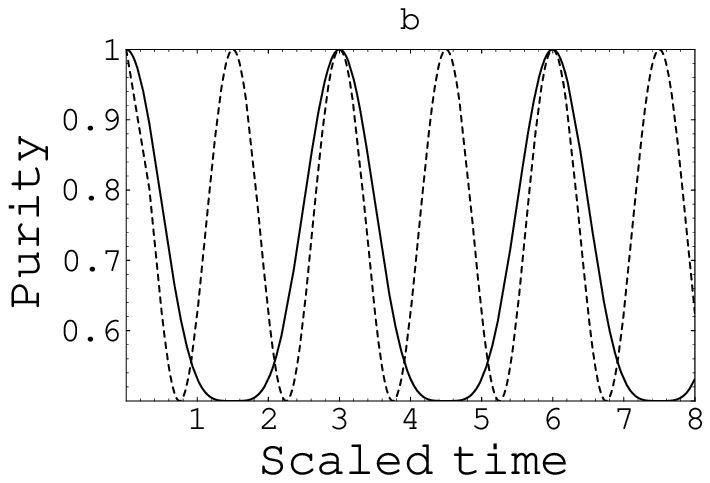}
 \includegraphics[width=15pc,height=9pc]{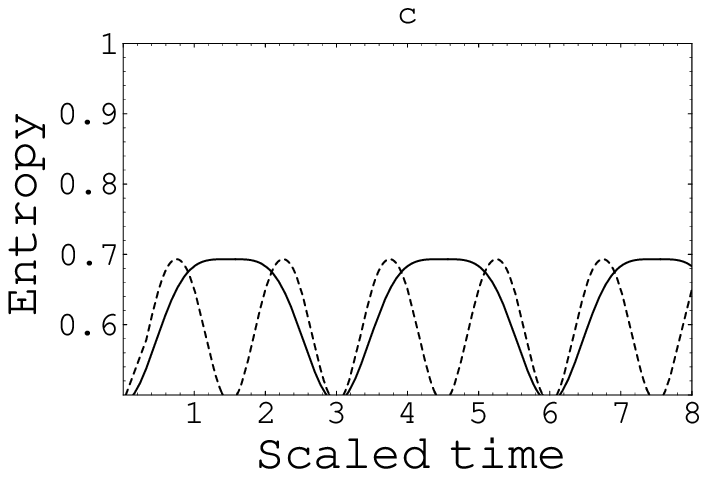}
  \caption{The same as Fig.$(1)$, with $t_x=-1$ and $t_y=t_z=0$.}
  \end{center}
\end{figure}

{\it Second class:} For this class, we assume that Alice and Bob
both have the same qubit but on a different polarization as:
\begin{equation}
\rho_a=\frac{1}{2}(1+\sigma_x),\quad \rho_b=\frac{1}{2}(1-\tau_x),
\end{equation}
Namely $s_x=1, t_x=-1$ and the other components vanish. After the
evolution, the joined density operator is defined by the Bloch
vectors
\begin{eqnarray}
\row{s}&=&(cos2t\theta_1\cos2t\theta_3-sin2t\theta_1\sin2t\theta_3,\quad
0,\quad 0), \nonumber\\
\row{t}&=&(\cos2t\theta_1(1-2\cos^2t\theta_3),\quad
-cos2t\theta_2(1-2\cos^2t\theta_3),\quad 0),
\end{eqnarray}
in addition to the dyadic $\dyadic{C}$, which is defined by
\begin{eqnarray}
c_{xx}&=&c_{xy}=c_{xz}=0, \nonumber\\
c_{yx}&=&c_{yy}=0,
c_{yz}=\sin2t\theta_1(1-2cos^22t\theta_3)-\cos2t\theta_1\sin2t\theta_3,
\nonumber\\
c_{zx}&=&0,c_{zy}=-\sin2t\theta_1(1-2cos^22t\theta_3)-\cos2t\theta_1\sin2t\theta_3,\quad
c_{zz}=0,
\end{eqnarray}
Also,  the individual qubits  are specified by their new Bloch
vectors
\begin{eqnarray}
\tilde{\row{s}}&=&(\cos2t\theta_1\cos2t\theta_3,\quad 0,\quad 0),
\nonumber\\
\tilde{\row{t}}&=&(-\cos2t\theta_1\cos2t\theta_3+\sin2t\theta_1\sin2t\theta_3,\quad
0,\quad 0).
\end{eqnarray}
In Fig.$(2)$, we plot the degree of entanglement against the scaled
time. From this figure, it is clear that the  generated state is
entangled most of the time except at specific points. At these
points the two qubits are completely separated. This is clear from
Fig.$(2b)$, where we plot the purity of the two qubits. We see when
the purities of the two qubits are completely swapped, the two
qubits become separable states. This occurs at normalized
$t=1.5,2.5,3.5$ etc. On the other hand, when the purities of the two
qubits  coincide, the generated entangled state has a large degree
of entanglement and it reaches its maximum value when the purities
are maximum. This phenomena is seen  once the interaction starts as
well as  at normalized  $t=3n, n=1,2,3,...$. This feature is
confirmed from Fig$(2c)$, where the entropies of the two qubits are
plotted. In some intervals neither Alice nor Bob know   any
information about the  other. At these instance all the information
has been transformed from one pair to each other. Alice can employ
this technique to send information to Bob.

{\it Third class:} In this class, we assume that Alice has the
same qubit as before, while Bob's qubit is polarized  along  $x$
and $y$ axis. In  explicit form
\begin{equation}
\rho_a=\frac{1}{2}(1+s_x\sigma_x),\quad
\rho_b=\frac{1}{2}(1+t_x\tau_x+t_y\tau_y).
\end{equation}
After the evolution the joined density operator $\rho_{ab}$  is
defined by  the following Bloch vectors
\begin{eqnarray}
\row{s}&=&(s_x\cos2t\theta_1\cos2t\theta_3+t_x\sin2t\theta_1\sin2t\sin\theta_3,
-t_y\sin2t\theta_1\sin\theta_3,\quad 0), \nonumber\\
\row{t}&=&\bigl(-t_x\cos2t\theta_1(1-2\cos^2t\theta_3)+s_x\cos2t\theta_1\sin2t\sin\theta_3,
\nonumber\\
 &&-t_x\cos2t\theta_2(1-2\cos^2\theta_3),\quad 0\bigr),
\end{eqnarray}
and the cross dyadic $\dyadic{C}$
\begin{eqnarray}
c_{xx}&=&c_{xy}=0,c_{xz}=-t_y\sin2t\theta_2(1-2\cos^2t\theta_3),
\nonumber\\
c_{yx}&=&c_{yy}=0,c_{yz}=-t_x\sin2t\theta_1(1-2\cos^2t\theta_3)-s_x\cos2t\theta_1\sin2t\theta_3,
\nonumber\\
c_{zx}&=&-t_y\cos2t\theta2\sin2t\theta_3,
 \nonumber\\
c_{zy}&=&-s_x\sin2t\theta_1(1-2\cos^2t\theta_3)+t_x\sin
2t\theta_3\cos2t\theta_1, \quad c_{zz}=0.
\end{eqnarray}
\begin{figure}
  \begin{center}
  \includegraphics[width=25pc,height=10pc]{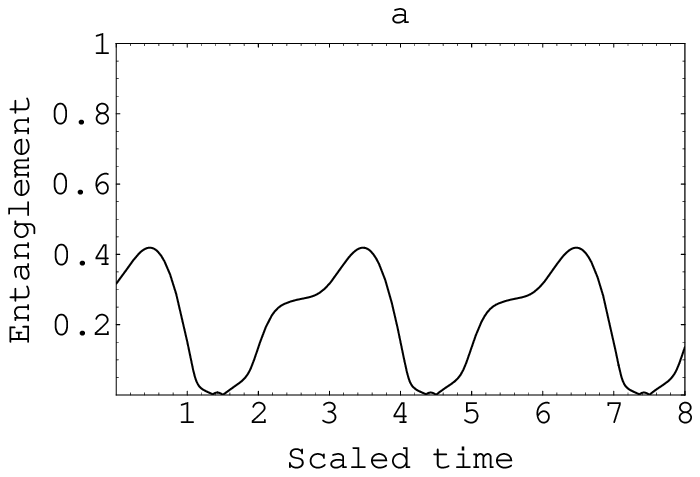}
 \includegraphics[width=15pc,height=9pc]{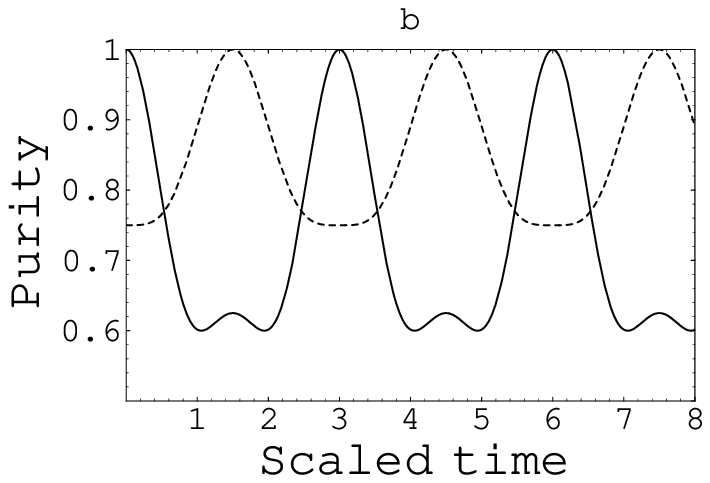}
 \includegraphics[width=15pc,height=9pc]{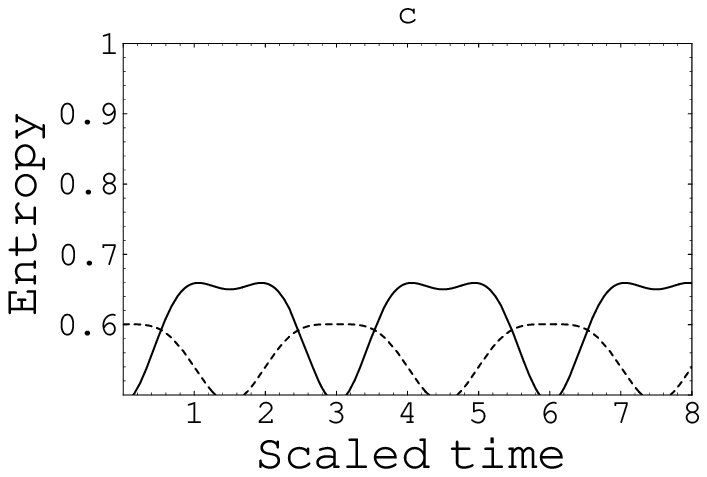}
  \caption{The same as Fig.$(1)$, with $t_x=t_y=0.5$ and  $t_z=0$.}
  \end{center}
\end{figure}
On the other hand  after the evolution, the two qubits evolve as
\begin{eqnarray}
\tilde\rho_a&=&\frac{1}{2}(1+s_x\cos2t\theta_1\cos2t\theta_3\sigma_x-t_y\sin2t\theta_3\sigma_y),
\nonumber\\
\tilde\rho_b&=&\frac{1}{2}\bigl(1+(t_x\cos2t\theta_1\cos2t\theta_3+s_x\sin2t\theta_1\sin2t\theta_3)\tau_x
\nonumber\\
&&\quad\quad -t_y\cos2t\theta_2\cos2t\theta_3\tau_y\bigr),
\end{eqnarray}

In Figs.$3a$,  we have plotted the amount of entanglement
contained in the generated state $\rho_{ab}$.  We consider $s_x=1$
while $t_x=t_y=0.5$, and  assume that Alice and Bob both perform
the same unitary operator with $\frac{\pi}{6}$. We can show that
the amount of entanglement is larger than the entanglement which
is depicted in the first class, but less than the  second class.
Also the purity of the initial state of Bob is larger than the
first class and smaller than the second  one. In Fig.$3c$, the
behavior of the entropies of the individual qubit is shown. This
figure depicts a partially swapping of the purity. This is clear
where the purity of Alice's qubit  never reaches   zero. It
decreases with amount enough for Bob's qubit to reach  the maximum
purity. The amount of information which is known by Alice and Bob
about the qubit of each other is plotted in Fig.$3c$. It is clear
that in this class, Bob is not able to  know all the information
contained in Alice's qubit.

\section{Entangled Class}
We consider a class of pure entangled states  defined by its joint
density operator
\begin{equation}
\rho_{ab}=\frac{1}{4}(1+p\sigma_x-p\tau_x-c_{xx}-q(c_{yy}+c_{zz})),
\end{equation}
where $q=\sqrt{1-p^2}$. This class of entangled state is very
important, where for $p=0$, one get the singlet state and
separable for $p=1$.
 This class has been studied in \cite{Nm}, where we
quantify the amount of exchange information between this state and
environment. Under the effect of the unitary operator, this state
is transformed such that,

\begin{eqnarray}\label{eq:41}
\row{s}&=&(p\cos2t(\theta_1+\theta_3),0,0), \nonumber\\
\row{t}&=&(p\cos2\theta_1(1-2\cos^22t\theta_3)+p\cos2t\theta_1\sin2t\theta_3,-p\cos2t\theta_2(1-2\cos^2t\theta_3),0),
\nonumber\\
\end{eqnarray}
 and
\begin{eqnarray}\label{eq:42}
c_{xx}&=&-1, c_{xy}=c_{xz}=0,
 \nonumber\\
c_{yx}&=&0,\quad c_{yy}=-1,\quad
c_{yz}=p\sin2t\theta_1(1-2\cos^2t\theta_3)-p\cos2t\theta_1\sin2t\theta_3,
\nonumber\\
c_{zx}&=&0,\quad c_{zy}=-p\cos2t(\theta_1+\theta_3),\quad
c_{zz}=-q.
\end{eqnarray}
On the other hand, if we trace out one qubit, we can obtain the
other, so one can obtain the density operators for Alice and Bob
qubis as
\begin{equation}
\tilde\rho_a=\frac{1}{2}(1+p\cos2t\theta_1\cos2t\theta_3\sigma_x),\quad
\tilde\rho_b=\frac{1}{2}(1-p\cos2t(\theta_1-\theta_3)\tau_x).
\end{equation}
\begin{figure}[htp]
  \begin{center}
  \includegraphics[width=25pc,height=10pc]{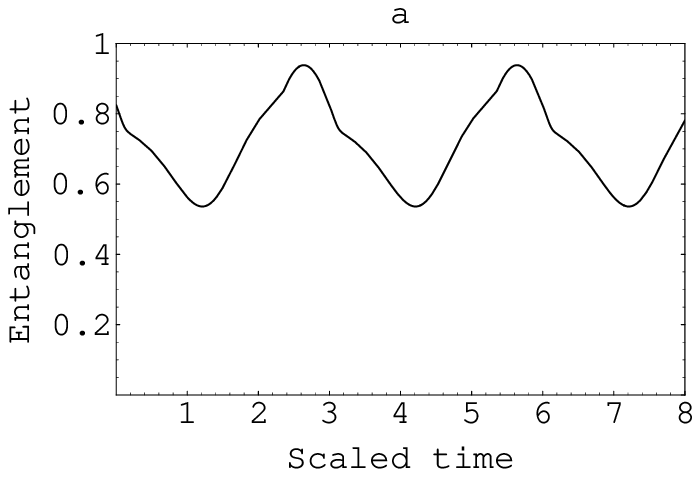}
 \includegraphics[width=15pc,height=9pc]{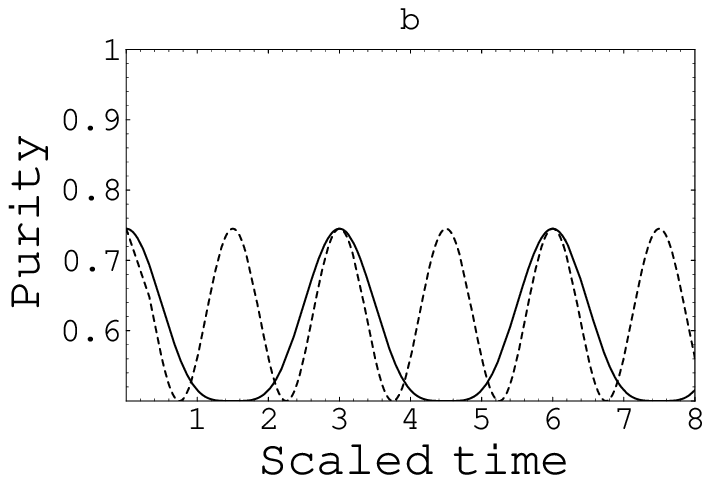}
 \includegraphics[width=15pc,height=9pc]{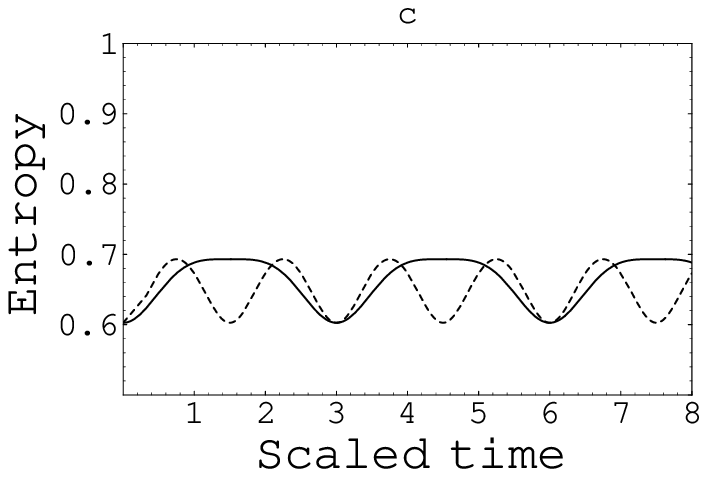}
  \caption{The figure is evaluated for $p=0.7$ and $q=\sqrt{1-p^2}$.
   (a) the degree of entanglement for $\rho_{ab}$,
after the evolution  (b) The purities for the Alice 's state
$\rho_a$( the solid line) and Bob'state $\rho_b$(dot -line). (c) The
Entropy for $\rho_a$ (solid line) and $\rho_b$ (dot line) }
  \end{center}
\end{figure}
The behavior of the amount of entanglement contained in the
entangled state, which is defined by (\ref{eq:41}) and
(\ref{eq:42}), is shown in Fig$(4a)$. We consider that the initial
state has $p=0.7$ as an initial state, where this parameter
completely defines the state. First of all the state is always
entangled and the degree of entanglement   increases and decrease
but never reach zero. In Fig.$4b$ and $4c$, we plot the behavior
of the purities and entropies of the individual systems,
$\tilde\rho_a$ and $\tilde\rho_b$, respectively. From these two
figures, we can see that the purity of one qubit may  be increased
on the expense of the other qubit.  On the other hand the entropy
of each of them can not reach zero. This means that, the two
qubits still always have some information about each other.

\section{ Conclusion}
In this contribution, we study the interaction of a two separable
qubits by means of  the canonical unitary operator. An analytical
expression of the joint density operator is obtained. We calculate
explicitly the reduced density operator for each qubit. Different
classes of initial density operators are considered. The amount of
entanglement contained in the joint entangled state is quantified
by using the negativity measurement. The swapping phenomena of
purity and  entropy is investigated for the reduced density
operator. It is clear that the unitary operator parameters
$\alpha_i,~ i=1,2,3$ play the central role in the interaction
process. If the state of one qubit is different from the other
qubit or polarized in an opposite direction, then the information
is completely transformed from one qubit to the other.

\end{document}